\documentclass{llncs}
\usepackage{graphicx, subfigure}
\usepackage{url}
\usepackage{textcomp}
\usepackage{amssymb,amsfonts,amsmath,graphicx,cite,csquotes}
\usepackage{subfigure,float}
\usepackage{color}
\usepackage{algorithm}
\usepackage{algorithmic}
\usepackage{authblk}
\begin{document}

%
\title{Health Monitoring of Critical Power System Equipments using Identifying Codes}
%
\author{Kaustav Basu$^1$, Malhar Padhee$^2$, Sohini Roy$^1$, Anamitra Pal$^2$, Arunabha Sen$^1$, Matthew Rhodes$^3$, and Brian Keel$^3$}

%
%
\institute{NetXT Lab, SCIDSE, Arizona State University \and 
                      Pal Lab, SECEE, Arizona State University \and 
                      Salt River Project \\
                      \email{kaustav.basu@asu.edu, Malhar.Padhee@asu.edu, Sohini.Roy@asu.edu, Anamitra.Pal@asu.edu, asen@asu.edu, Matthew.Rhodes@srpnet.com, Brian.Keel@srpnet.com}}

\maketitle              
\vspace{-18.00pt}
\begin{abstract}
High voltage power transformers are one of the most critical equipments in the electric power grid. A sudden failure of a power transformer can significantly disrupt bulk power delivery. Before a transformer reaches its critical failure state, there are indicators which, if monitored periodically, can alert an operator that the transformer is heading towards a failure. One of the indicators is the signal to noise ratio (SNR) of the voltage and current signals in substations located in the vicinity of the transformer. During normal operations, the width of the SNR band is small. However, when the transformer heads towards a failure, the widths of the bands increase, reaching their maximum just before the failure actually occurs. This change in width of the SNR can be observed by \textit{sensors}, such as phasor measurement units (PMUs) located nearby. Identifying Code is a mathematical tool that enables one to \emph{uniquely} identify one or more {\em objects of interest}, by generating a \emph{unique signature} corresponding to those objects, which can then be detected by a sensor. In this paper, we first describe how Identifying Code can be utilized for detecting failure of power transformers. Then, we apply this technique to determine the fewest number of sensors needed to \emph{uniquely} identify failing transformers in different test systems. 
\end{abstract}
\vspace{-16.00pt}
\section{Introduction}
The electric power grid is arguably the most critical of all the infrastructures as other infrastructures, such as, communication, transportation and finance are heavily dependent on it. Similarly, high voltage (HV) power transformers, generators, and transmission lines are the most critical components of the electric power grid. 
Therefore, an untimely loss of HV transformers can be catastrophic for not only the electrical infrastructure, but also the other critical infrastructures that depend on it. Accordingly, it will be helpful if it can be recognized before the event, that a transformer is heading towards a failure, so that corrective measures can be undertaken. Fortunately, before a transformer reaches its critical failure state, there are ``cues''(or indicators) which, if monitored periodically, can alert an operator that the transformer is heading towards a failure. One of the indicators is the signal to noise ratio (SNR) of the voltage and current signals in substations located in the vicinity of the transformer. During normal operations, the width of the SNR bands are small. However, when the transformer heads towards a failure, the widths of the bands increase, reaching their maximum just before the failure actually occurs. This change in width of the SNR can be observed by phasor measurement units (PMUs) located nearby. 

Identifying Code is a mathematical tool that enables one to \emph{uniquely} identify one or more {\em objects of interest}, by generating a \emph{unique signature} corresponding to those objects, which can then be detected by a sensor. In this paper, the objects of interest are HV transformers. When a transformer is heading towards failure, it generates ``indicators'', which, if monitored by some ``sensors'', may provide information to an operator in the control center about the impending failure of the transformer. Since the number of transformers in the grid is large, and the sensors are expensive, one would like to deploy as few sensors as possible (fewer than the number of transformers) and yet retain the capability that, when a transformer is heading towards a failure, it can be \emph{uniquely} identified. 

PMU is a device that can be utilized as a ``sensor'' for monitoring the health of transformers. When placed on a generator, load, or zero injection bus, in the power grid, PMUs give the voltage of that particular bus, as well as the currents flowing in the branches (lines  or transformers) incident on that bus (while being subjected to the PMU\textquotesingle s measurement channel limitations). Since a power transformer can \emph{only} be placed between two buses, a judicious placement of a few PMUs (sensors) can effectively monitor health of all the transformers, and in case a transformer heads towards a failure, the sensors can create a \emph{unique fault signature} that enables the operator to identify the troubled transformer.



In this paper, we, (i) describe the Rudd power transformer failure incident that motivated this study, (ii) describe how Identifying Code can be utilized for \emph{unique} identification of the transformers that are heading towards a failure, and, (iii) provide a technique to compute the fewest number of sensors to be deployed, to ensure \emph{unique} identification of the transformers that are heading towards a failure in standard test systems.



\section{Related Work}
\label{Related Work}
Prior research on health monitoring using PMUs have been mostly directed towards improving security and stability of the power system \cite{R38}. In addition, a number of studies have focused on placement of PMUs \cite{R10, R11} to realize a variety of objectives. The problem under study in this paper can also be viewed as a PMU placement problem as it computes the fewest number of PMUs and their locations, so that the \emph{unique} identification capability is realized. It is important to highlight here that none of the PMU placement strategies proposed so far had the unique identification capability as the objective for PMU deployment. 


Karpovsky {\em et. al.} introduced the concept of Identifying Codes in  \cite{R24} and  provided results for Identifying Codes for graphs with specific topologies, such as binary cubes and trees. 
Using Identifying Codes, Laifenfeld {\em et. al.} studied covering problems in \cite{R25}. A special case, where only a subset of nodes needs a unique code, can be modeled with a bipartite graph, and was studied as ``Discriminating Codes'' in \cite{Discriminating}. This special case is relevant for our study as we focus on finding unique signatures for a subset of nodes, instead of all the nodes, as is done in Identifying Codes.




\section{Lessons Learnt from Rudd Power Transformer Failure}

During the early hours of June 1, 2016, a large power transformer at the Rudd substation of Salt River Project (SRP), a large utility company in Arizona, suddenly caught fire. A 27,000-gallon tank of mineral oil used as a transformer coolant, burned, and spewed thick smoke over a large area. A few snapshots are illustrated in Figs. 1(a) and 1(b) \cite{R16}. The cause of the failure was identified to be bushing failure. Due to the redundancy present in the system design as well as the fact that the fire broke out during low-load conditions (system load is small in early morning), no power outages occurred. This incident highlights the need for better monitoring techniques for these critical and expensive equipments.
\vspace{-12.00pt}
\begin{figure*}[thb]
	\begin{center}
		\centering
		\subfigure[]{\label{footprint}\includegraphics[width=0.40\textwidth, keepaspectratio]{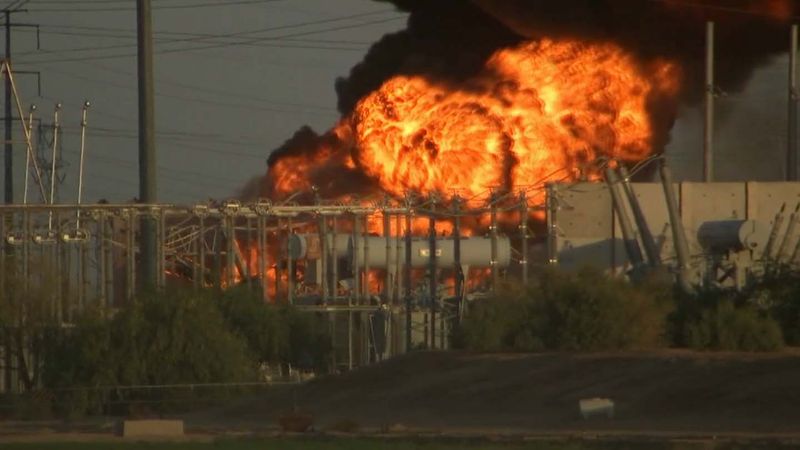}}
		\subfigure[]{\label{constellation}\includegraphics[width=0.40\textwidth, keepaspectratio]{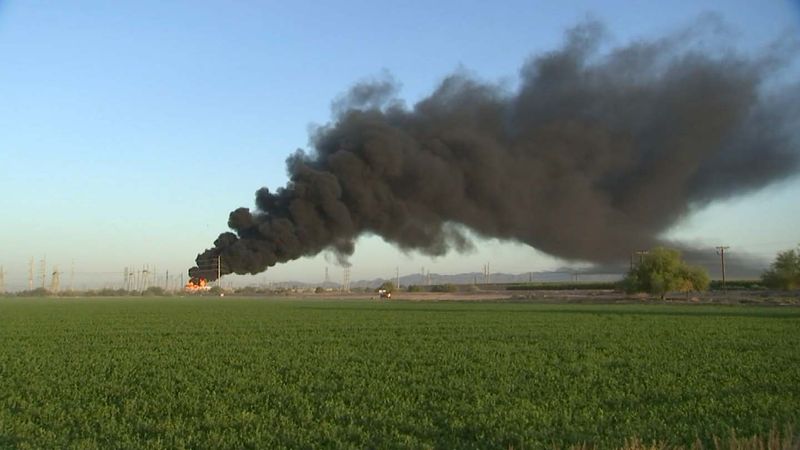}}
		
	\end{center}
	\vspace{-8.0mm}
	\caption{Transformer fire at Salt River Project (SRP)'s Rudd substation in Avondale \cite{R16}.}
	\vspace{-5.0mm}
\end{figure*}


SRP shared their operational data leading up to the failure of this transformer with us for analysis.
Because causes of such failures gradually build-up over time, if one is paying attention, the signs of an impending failure may be observable ``days'' before the actual failure event. PMUs continuously produce outputs at a very fast rate (typically 30 samples per second). When placed near transformers, PMUs, through their measurements, can serve as {\em sensors} to monitor the health of the transformer, and capture degradation in the health of a transformer over time. 
It may be noted that a PMU provides complex voltage and current measurements at the bus where it is placed. If the PMU has to serve as a sensor for monitoring transformer health, it must have a way to measure it with a ``cue'' (or indicator or metric). This metric should be independent of the ``unit" of the measured quantity (either voltage or current), so that a proper comparison can be made. Signal to noise ratio (SNR), a classical measure of the quality of a signal, can serve as this desired metric. It compares the level of a signal to the level of background noise that is present in it. Mathematically, the SNR of a signal can be expressed as reciprocal of the coefficient of variation, i.e., the ratio of its mean to its standard deviation, as shown in Eq. \ref{eq1}.

\begin{equation}
\label{eq1}
SNR \ (in \ dB) = 10*log{\frac\mu \sigma}
\end{equation}

\begin{figure*}[h!]
	\begin{center}
		\centering
		\subfigure[]{\label{footprint}\includegraphics[width=0.85\textwidth, keepaspectratio]{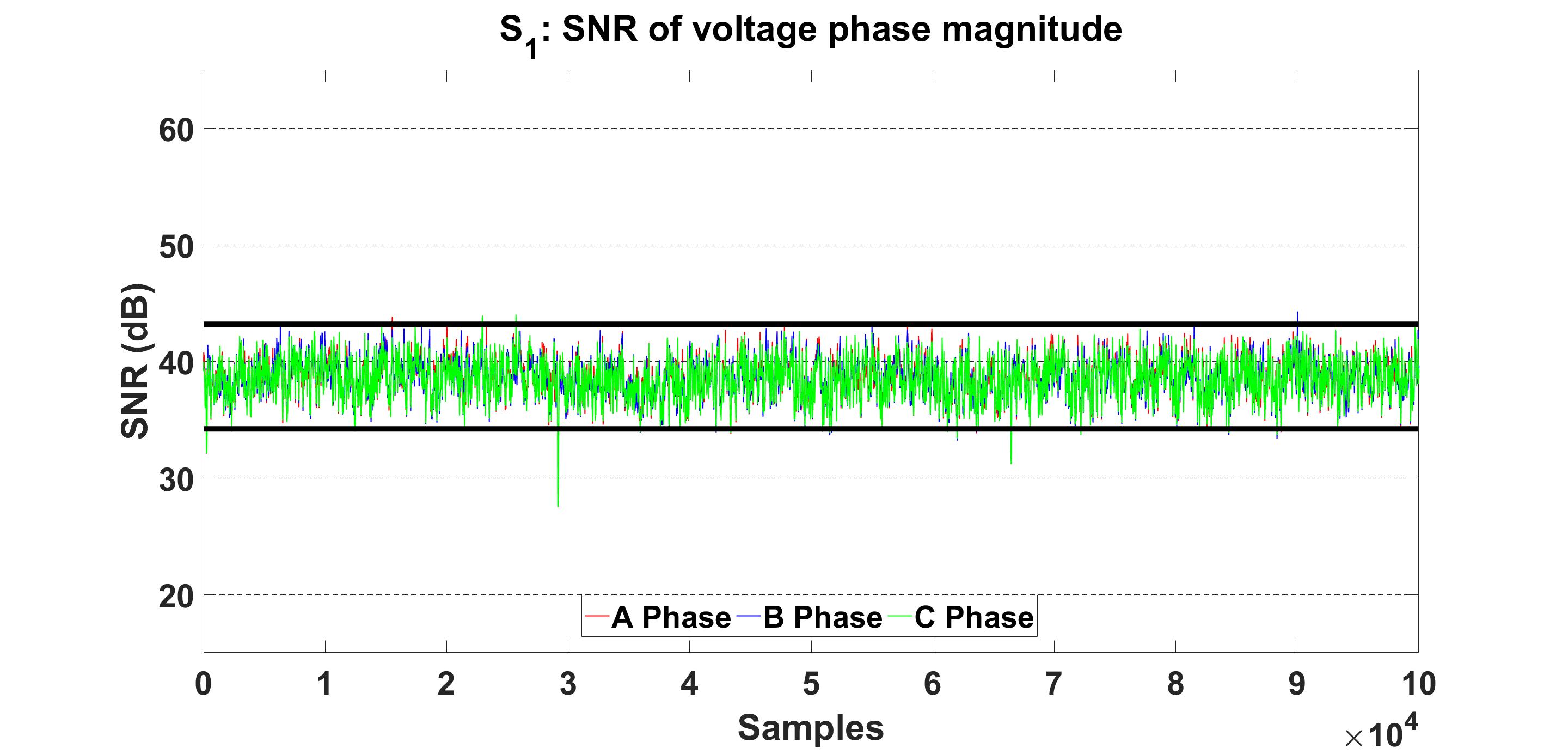}}
		\subfigure[]{\label{constellation}\includegraphics[width=0.85\textwidth, keepaspectratio]{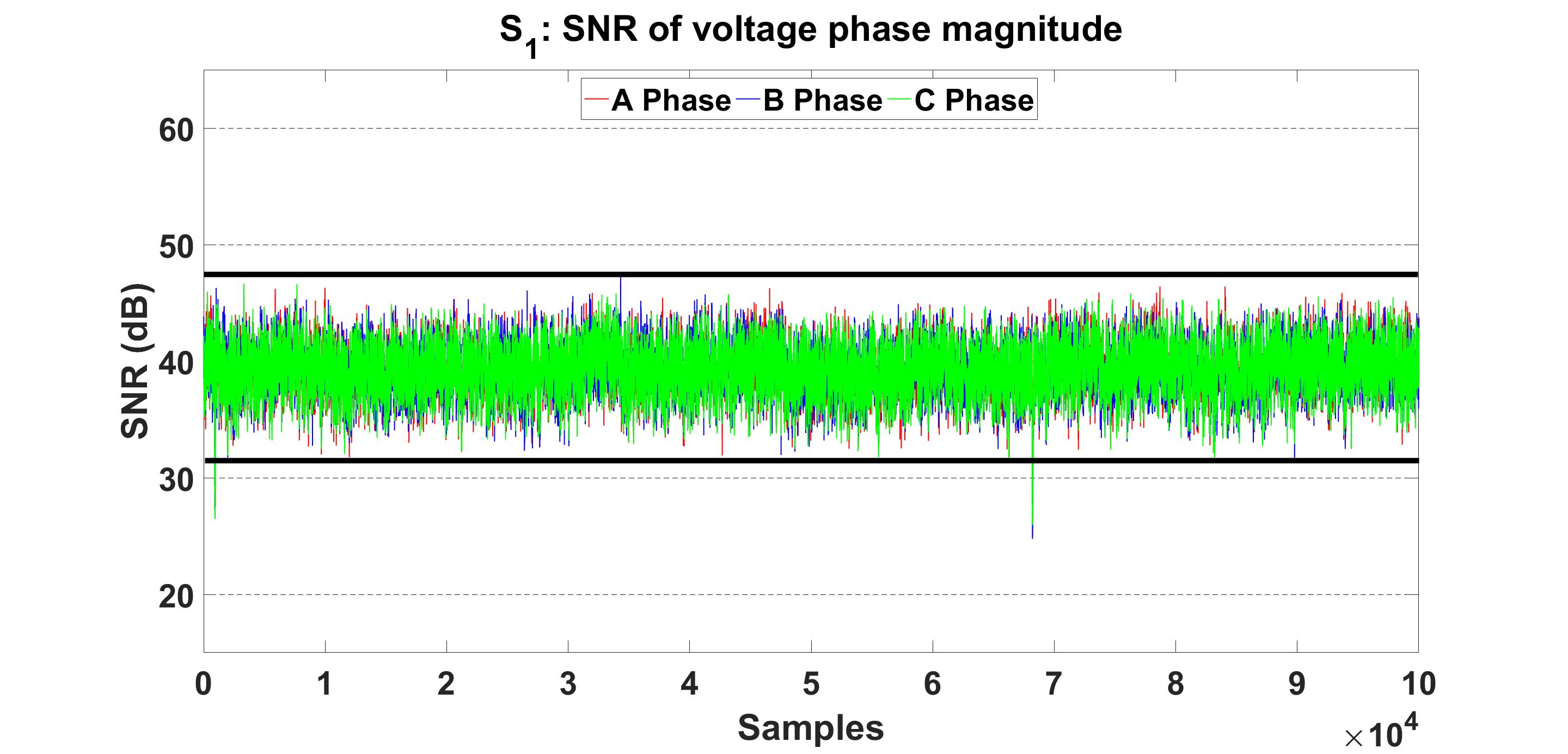}}
		\subfigure[]{\label{footprint}\includegraphics[width=0.85\textwidth, keepaspectratio]{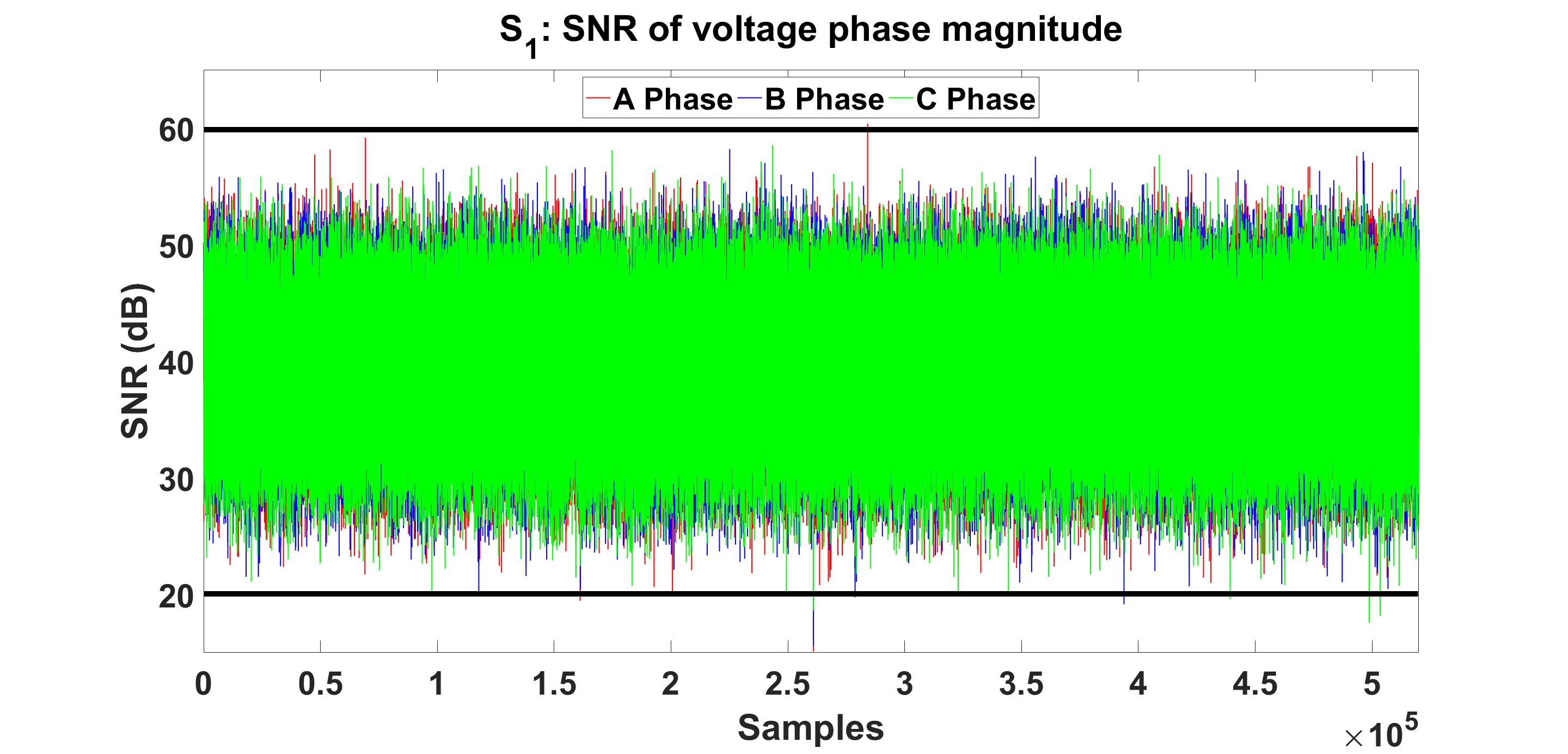}}
	\end{center}
	\vspace{-8.0mm}
	\caption{Variation in width of SNR as one moves closer (in time) to instant of failure.}
    \label{width}
\end{figure*}

\begin{figure*}[tbh]
	\begin{center}
		\subfigure[]{\label{constellation}\includegraphics[width=0.85\textwidth, keepaspectratio]{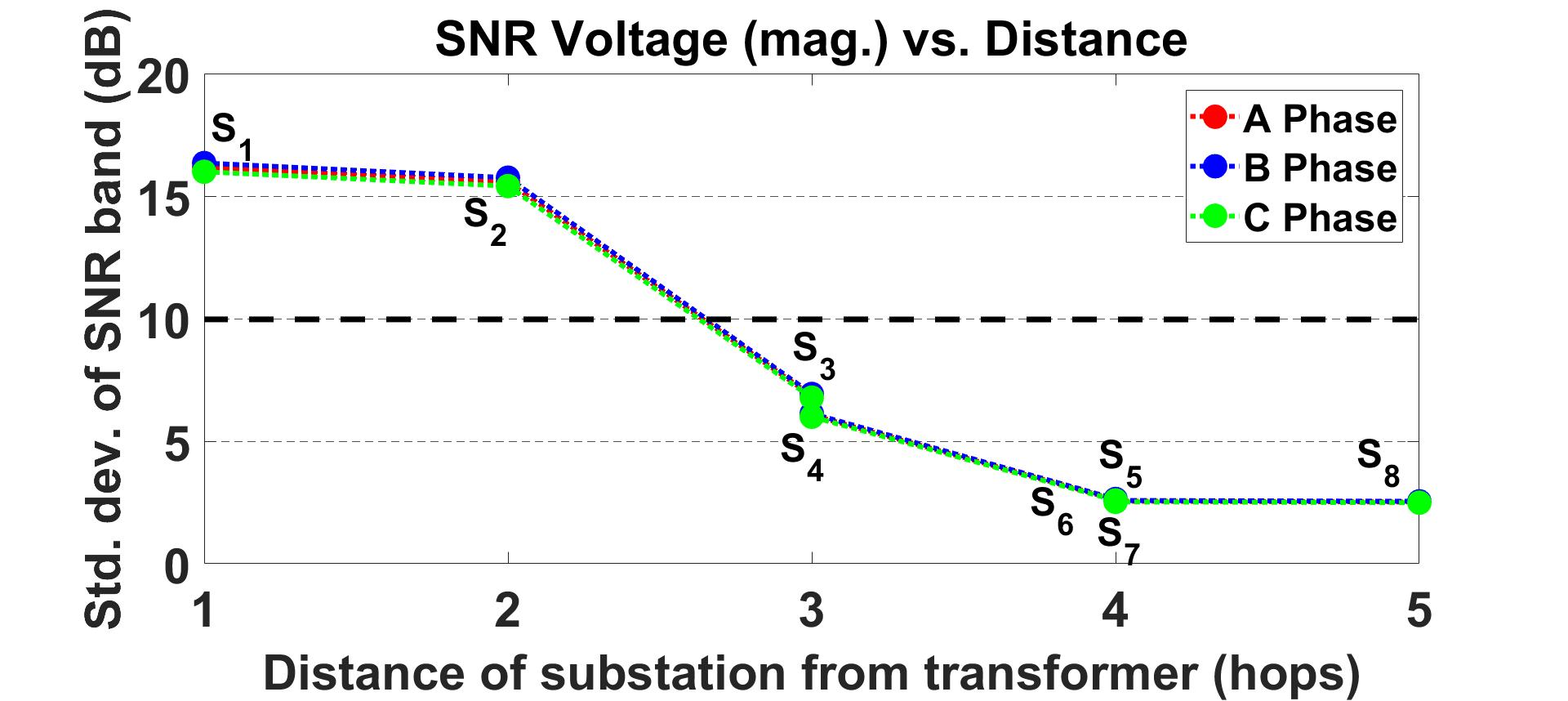}}
	\end{center}
	\vspace{-8.0mm}
	\caption{Standard deviation of width of SNR as one moves (spatially) away from the failing equipment.}
    \label{distance}
    \vspace{-6.00mm}
\end{figure*}



\vspace{-0.00pt}
In Eq. \ref{eq1}, $\mu$ is the signal mean or expected value and $\sigma$ is the standard deviation, or an estimate thereof. It is difficult to directly compare different signals (such as voltages and currents). However, SNR (in decibels) is a relative metric and therefore, it can be used to compare diverse signals and create alerts/alarms. The Rudd transformer failure data obtained from SRP, comprises of PMU readings (voltages and currents) one year away from the day of the failure (June 1, 2016) up to the data collected only a few hours prior to the actual failure event. 

Two important pieces of observation were made from the SRP data. \\
\textit{Observation 1:} A steady growth in the width of the SNR bands (computed from the voltage magnitude measurements obtained from neighboring substations), was observed over a period of time, till the transformer failed. The observations for three instances of time, as it approached the actual time of failure, are shown in Fig. \ref{width}. 
Since the growth was similar in all three phases, it was concluded that the SNRs were capturing an event that was affecting all three phases, and not due to a single phase failure event, contributed by a current or a potential transformer failure. Moreover, as the width was uniform over the observed time period (an hour worth of data), it is clear that the captured event was \emph{not} a random transient event. \\
\textit{Observation 2:} In observation 1, we noted that the width of the SNR band at a specific PMU (sensor) location, increases as time approaches the actual failure event. From the data it was also clear that, as the distance of the PMU (sensor/monitoring device), from the transformer (monitored device) increased, the width of the observed SNR decreased. Fig. \ref{distance} shows the decrease in the width of the SNR bands as a function of the electrical distance (termed as hops) from the Rudd transformer. The data was collected from eight substations (S1, ..., S8) that neighbor Rudd, and had PMUs placed on them. It may be noted that the Rudd substation itself did not have a PMU on it during the time of failure.



Given that the deteriorating condition of a transformer can be noticed by PMUs located within a certain distance of the transformer, signals indicating the deteriorating condition, can be utilized to deploy effective monitoring strategies, so that an alarm is generated \emph{before} a transformer reaches a critical failure state. \emph{Identifying Code} is a mathematical tool that can be used for monitoring transformers in the power grid. Using this technique, the fewest number of sensors needed to enable an operator to \emph{uniquely} identify the failing transformer before it reaches a critical failure state can be computed.

\section{Overview of Identifying and Discriminating Codes}
\label{Identifying Code}
\begin{figure*}[h!]
	\begin{center}
		\includegraphics[width=0.60\textwidth, keepaspectratio]{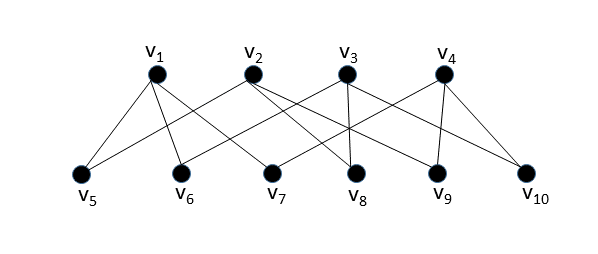}
		\vspace{-8.0mm}
        \caption{Graph with Identifying Code Set \{$v_1$, $v_2$, $v_3$, $v_4$\}}
		{\label{ICExample}}
        
	\end{center}
\end{figure*}

The notion of {\em Identifying Codes} \cite{R24} has been established as a useful concept for optimizing sensor deployment in multiple domains. In this paper, we use Identifying Code of the {\em simplest form} and define it as follows. {\em A vertex set $V'$ of a graph $G = (V, E)$ is defined as the Identifying Code Set (ICS) for the vertex set $V$, if for all $v \in V$, $N^+(v) \cap V'$ is unique where, $N^+(v) = v \cup N(v)$ and $N(v)$ represents the set of nodes adjacent to $v$ in $G = (V, E)$.} The {\em Minimum Identifying Code Set} (MICS) problem is to find the Identifying Code Set of {\em smallest cardinality}. The vertices of the set $V'$ may be viewed as {\em alphabets} of the code, and the {\em string} made up with the alphabets of $N^+(v)$ may be viewed as the unique ``code'' for the node $v$. For instance,
consider the graph $G = (V, E)$ shown in Fig. \ref{ICExample}. In this graph $V' = \{v_1, v_2, v_3, v_4\}$ is an ICS as it can be seen from Table \ref{exampletable} that $N^+(v) \cap V'$ is {\em unique} for all $v_i \in V$. From the table, it can be seen that the code for node $v_1$ is $v_1$, the code for $v_5$ is $v_1,v_2$, the code for $v_{10}$ is $v_3,v_4$, etc.
\vspace{-26.00pt}
\vspace{-30.00pt}
\begin{small}
	\begin{table}[h!]
		\centering
		\caption{$N^+(v) \cap V'$ results for all $v \in V$ for the graph in Fig. \ref{ICExample}}
		\begin{tabular} {| c  | c |}  \hline 
			    $N^+(v_1) \cap V' = \{v_1\}$ & $N^+(v_2) \cap V' = \{v_2\}$ \\ \hline
			    $N^+(v_3) \cap V' = \{v_3\}$ & $N^+(v_4) \cap V' = \{v_4\}$ \\ \hline
			    $N^+(v_5) \cap V' = \{v_1, v_2\}$ & $N^+(v_6) \cap V' = \{v_1, v_3\}$ \\ \hline
			    $N^+(v_7) \cap V' = \{v_1, v_4\}$ & $N^+(v_8) \cap V' = \{v_2, v_3\}$ \\ \hline
			    $N^+(v_9) \cap V' = \{v_2, v_4\}$ & $N^+(v_{10}) \cap V' = \{v_3, v_4\}$ \\ \hline
		\end{tabular}
		\label{exampletable}
		\vspace{-8.00mm}
	\end{table}
\end{small}

\noindent{\tt Graph Coloring with Seepage (GCS) Problem}: The MICS computation problem can be viewed as a novel variation of the classical Graph Coloring problem. We will refer to this version as the {\em Graph Coloring with Seepage (GCS)} problem. In the classical graph coloring problem, when a color is {\em assigned} (or injected) to a node, only that node is colored. The goal of the classical graph coloring problem is to use as few distinct colors as possible such that (i) every node receives a color, and (ii) no two adjacent nodes of the graph have the same color. In the GCS problem, when a color is assigned (or injected) to a node, not only does that node receive the color, but also the color {\em seeps} into all the adjoining nodes. For example, if a node $v_i$ is adjacent to two other nodes  $v_j$ and  $v_k$ in the graph, then if the color red is injected to $v_j$, not only $v_j$ will become red, but also $v_i$ will become red as it is adjacent to $v_j$. Now if the color blue is injected to $v_k$, not only $v_k$ will become blue, but also the color blue will seep in to $v_i$ as it is adjacent to $v_k$. Since $v_i$ was already colored red (due to seepage from $v_j$), after color seepage from $v_k$, it's color will be a {\em combination of red and blue (purple)}.  At this point, all three nodes  $v_j$, $v_k$, and $v_i$ will have ``distinct" colors red, blue, and purple, respectively. {\em The color assigned to a node may be due to: (i) only injection at that node, (ii) only seepage from other adjoining nodes and (iii) a combination of injection and seepage}. The colors injected at the nodes are referred to as {\em atomic} colors. The colors formed by the combination of two or more atomic colors are referred to as {\em composite} colors. The colors injected at the nodes (atomic colors) are all {\em unique}. The goal of the GCS problem is to inject colors to as few nodes as possible, such that (i) every node receives a color, and (ii) no two nodes of the graph have the same color.

\vspace{-2.00pt}
Suppose that the node set $V'$ is an ICS of a graph $G = (V, E)$ and $|V'| = p$.  In this case if $p$ distinct colors are injected to $V'$ (one distinct atomic color to one node of $V'$ ), then by the definition of ICS for all $v \in V$, if $N^+(v) \cap V'$ is unique, all nodes of  $G = (V, E)$ will have a unique color (either atomic or composite). Thus computation of MICS is equivalent to solving the GCS problem.

Identifying Code is useful when the goal is to monitor all nodes of the graph (i.e., each node is required to have a unique signature). However, in this paper our focus is on monitoring the health of only power transformers.
Moreover, in Identifying Code a color can be injected at any node of the graph (i.e., a sensor can be placed at any node of the graph). However, in the health monitoring problem, a sensor placed far away from the equipment to be monitored, may not be useful as ``cues'' (signals) indicating failing state of the equipment, may not even reach this sensor because of its distance from the equipment. Accordingly, some modification to the original concept of \emph{Identification} is needed. The following modifications are sufficient to capture the new scenario: (i) We identify a subset $V' \subseteq V$ that needs to receive a unique color; (ii) For each node $v \in V'$, we compute $N^k(v)$, where $N^k(v)$ represents the $k$-hop neighbors of $v$ (i.e., the set of nodes in the graph whose shortest path distance to $v$ is at most $k$); (iii). We construct a Bipartite graph $G' = (V_1 \cup V_2,  E)$ such that (a) $V_1 = V'$, (b) $V_2 = \cup_{v \in V_1}$ $N^k(v)$, and (iii) for nodes $v_i \in V_1$ and $v_j \in V_2$, there is an edge $e \in E$, if and only if $v_j \in N^k(v_i)$. With this modification, the transformer health monitoring problem with the fewest number of sensors is equivalent to computation of the smallest subset $V'_2 \in V_2$ such that injection of colors to this set of nodes ensures that each node in $V_1$ receives a unique color through {\em seepage}. In this study, we restrict our attention to $k = 1$ or $k = 2$ only, as cues of deteriorating health of transformer may not be observable at distances $k \geq 3$ (See Fig. \ref{distance}).

A variation of Identification Code when restricted to Bipartite graphs is known as {\em Discriminating Code} \cite{Discriminating}, and is defined as follows:  Let $G = (V_1 \cup V_2, E)$ be an undirected bipartite graph and let $N(v)$, denote the neighborhood of $v$, for any $v \in V_2$,  a subset $V'_2 \subseteq V_2$ is called the Discriminating Code of $G$ if $\forall v \in V_1, N(v) \cap V'_2$ is unique. We will refer to 
critical equipment health monitoring problem, with the fewest number of sensors, as the \emph{Monitoring Critical Equipment} (MCE) problem, which may be stated formally in the following way:

\vspace{0.045in}
\noindent
\textbf{MCE Problem}: Find the smallest subset $V'_2 \subseteq V_2$, such that injection of colors at these nodes, ensures that each node $v \in V_1$, receives a unique color through seepage. 


\section{Problem Formulation}
\label{Problem Formulation}

In this section, we formalize the problem of computing the fewest number of sensors to be deployed to monitor all critical equipments (HV transformers) in the power grid, so that, if they show signs of potential failure, then an operator in the control room, can uniquely identify them. Once the failing equipment is identified, corrective measures can be undertaken, such as a planned shutdown.  

\begin{figure*}[t!]
	\begin{center}
		\includegraphics[width=0.70\textwidth, keepaspectratio]{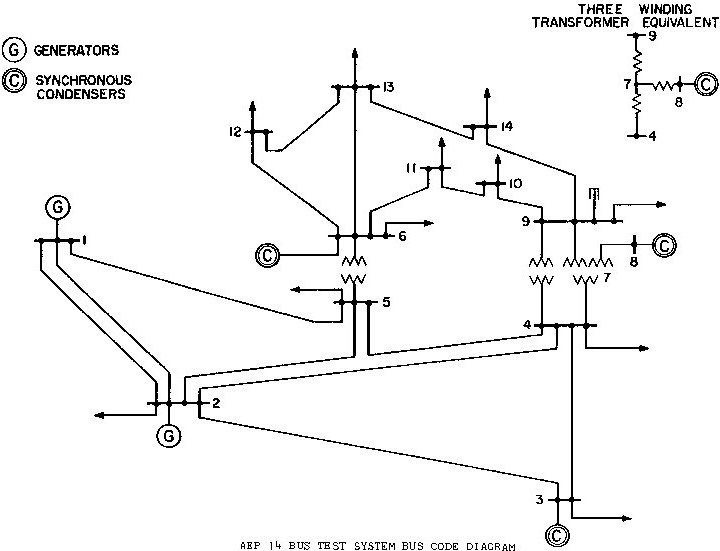}
        \caption{IEEE 14 Bus Test System}
		{\label{WSCC}}
        \vspace{-8.00mm}
	\end{center}
\end{figure*}

\begin{figure*}[h]
	\begin{center}
		\includegraphics[width=0.70\textwidth, keepaspectratio]{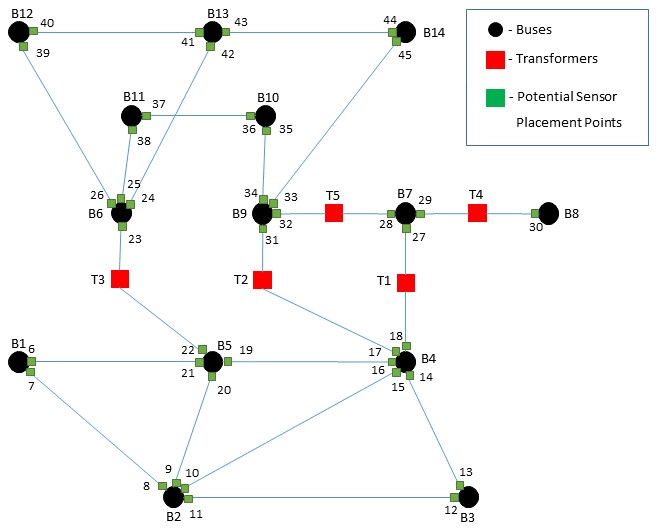}
        \caption{Potential Sensor Placement Locations in IEEE 14 Bus System}
		{\label{Potential}}
        \vspace{-8.00mm}
	\end{center}
\end{figure*}

\begin{figure*}[h!]
	\begin{center}
		\includegraphics[width=1.00\textwidth, keepaspectratio]{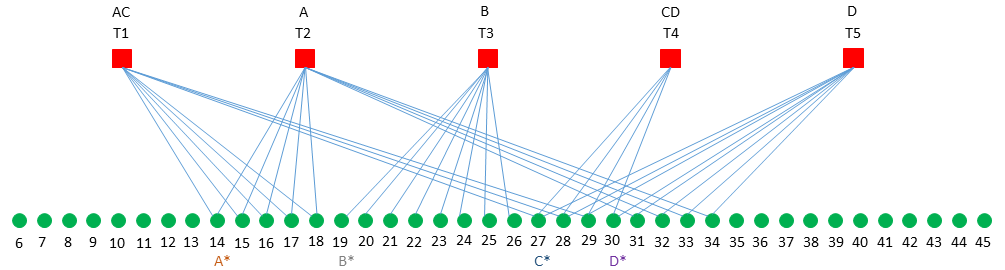}
        \caption{Bipartite Graph corresponding to IEEE 14 bus system with for k = 1}
		{\label{Bipartite1}}
        \vspace{-2.00mm}
	\end{center}
\end{figure*}

\begin{figure*}[h!]
	\begin{center}
		\includegraphics[width=1.00\textwidth, keepaspectratio]{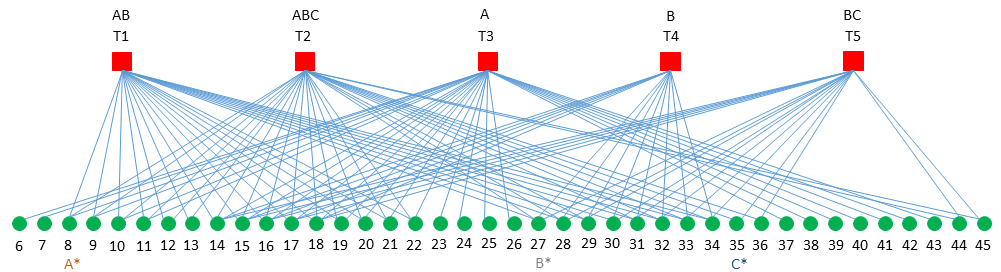}
        \caption{Bipartite Graph corresponding to IEEE 14 bus system with for k = 2}
		{\label{Bipartite2}}
        \vspace{-2.00mm}
	\end{center}
\end{figure*}

From our discussion in Section \ref{Identifying Code}, it is clear that Identifying Code relates to an underlying graph. In order to use Identifying Code to find the fewest number of sensors to be deployed to monitor critical equipments, we first have to construct a graph from the single line diagram (SLD) of the power system. Consider the IEEE 14 Bus System shown in Fig. \ref{WSCC}. We construct a graph $G = (V, E)$ from the SLD, where each node represents either a bus or a transformer, and two nodes are connected by an edge if the corresponding buses, or bus and transformer are connected. The Fig. \ref{Potential} shows the graph $G = (V, E)$ constructed from the IEEE 14 Bus SLD, shown in Fig. \ref{WSCC}. In Fig. \ref{Potential}, the buses are represented by black circular nodes and the transformers by red square nodes. In power systems, the monitoring devices (such as the PMUs) can be placed on the ends of the transmission lines, next to the buses \cite{R10}. In Fig. \ref{Potential}, the potential locations where a monitoring device can be deployed are shown by small green squares. \\
The goal of this exercise is to determine the health of the red squares (transformers) before they reach a critical state. Signal of failing health of a red square reaches only up to a certain distance from the location of the red square, where distance is measured in terms of number of hops. The monitoring devices can only be placed at the green squares. If we assume that the signal of failing health of a red square can reach $k$ hops, then all green squares within $k$ hop distance of the red square will recognize that that particular red square (transformer) is failing. This can be captured in a bipartite graph $G = (V_1 \cup V_2, E)$, where each node $v \in V_1$ represents a red square and each node $v \in V_2$ represents a green square. There is an edge $e \in E$ connecting nodes $v_i \in V_1$ and $v_j \in V_2$ if the signal from the red square $r_i$, represented by node $v_i$ in Fig. \ref{Potential}, can reach the green square $g_j$, represented by node $v_j$ in Fig. \ref{Potential}. Such graphs corresponding to the IEEE 14 Bus System, with $k = 1$ and $k = 2$, are shown in Fig. \ref{Bipartite1} and Fig. \ref{Bipartite2}, respectively. Since the IEEE 14 Bus System has 5 transformers (red squares in Fig. \ref{Potential}), the vertex set $V_1$ in the bipartite graphs shown in Figs. \ref{Bipartite1} and \ref{Bipartite2} has 5 nodes. Since, in the IEEE 14 Bus System, there are 40 potential locations for placement of sensors (green squares), in Fig. \ref{Potential}, the vertex set $V_2$, in the bipartite graphs shown in Figs. \ref{Bipartite1} and \ref{Bipartite2}, has 40 nodes (numbered from 6-45), denoted by green circles. It may be noted that when $k = 1$, only 21 out of 40 potential locations are viable locations for placement of sensors as the other 19 locations are not within 1-hop neighborhood of the transformers. However, when $k = 2$, all 40 nodes are viable locations for placement of sensors, as all of them are within the 2-hop neighborhood of the transformers. It may be noted that some of the nodes in Figs. \ref{Bipartite1} and \ref{Bipartite2}, are labeled with strings such as ``A'', ``AC'', etc. The explanation and significance of these strings are given in Section. \ref{Experimental Results}.

\vspace{-4.00pt}

\section{Problem Solution}
\label{Problem Solution}
\vspace{-2.00pt}
In this section, we provide an Integer Linear Programming (ILP) formulation for solving the MCE problem, as stated below. 

\noindent{\textit{Instance:}} $G = (V_1 \cup V_2, E)$, an undirected bipartite graph.\\
\textit{Problem}: Find the smallest subset $V'_2 \subseteq V_2$, such that injection of colors at these nodes, ensures that each node $v_i \in V_1$, receives a unique color (either atomic or composite) through seepage.

\noindent {We use the notation $N(v_i)$ to denote the neighborhood of $v_i$, for any $v_i \in V_1 \cup V_2$.} Corresponding to each $v_i \in V_2$, we use an indicator variable $x_i$,
\[x_i = \left\{ \begin{array}{ll}
                    1, & \mbox{if a color is injected at node $v_i$, } \\
                    0, & \mbox{otherwise}
                    \end{array}
            \right. \]

\begin{center}
\begin{tabular}{ l l l } 
 
 \textit{Objective Function:} & \ \textit{Minimize} \ $\sum_{v_i \in V_2}x_i$ &  \\ \\ 
 \textit{Coloring Constraint:} & \ $\sum_{v_i \in N(v_j)}x_i \geq 1,$ & \ $\forall v_j \in V_1$ \\ \\
 \textit{Unique Coloring Constraint:} & \ $\sum_{v_i \in \{N(v_j) \bigoplus N(v_k)\} }x_i \geq 1,$ & \ $\forall v_j \neq v_k, \in V_1$ \\ \\

\end{tabular}
\end{center}

\noindent{$N(v_j) \bigoplus N(v_k)$ denotes the Exclusive-OR (symmetric set difference) of the node sets $N(v_j)$ and $N(v_k)$.}
It may be noted that the objective function ensures that the fewest number of nodes in $V_2$ are assigned a color. The Coloring Constraint ensures that every node in $V_1$ receives at least one color through seepage from the colors injected at nodes in $V_2$. A consequence of the Coloring Constraint is that, a node in $V_1$ may receive more than one color through seepage from the colors injected at nodes in $V_2$. The Unique Coloring Constraint ensures that, for every pair of nodes ($v_j, v_k$) in $V_1$, at least one node in the node set $N(v_j) \bigoplus N(v_k) \subseteq V_2$ is injected with a color. This guarantees that $v_j$ and $v_k$ will not receive identical colors through the color seepage from the nodes in $V_2$.  
\vspace{-8.00pt}
\section{Experimental Results and Discussion}
\label{Experimental Results}

In this section, we present the results of of our technique on standard power system test cases, such as IEEE 14, 30, 57, 118, PEGASE 89 bus, and Polish 2383 bus systems. As discussed in Section. \ref{Problem Formulation}, the IEEE 14 bus system has 5 transformers and 40 potential locations for placement of sensors. The bipartite graphs for the IEEE 14 bus system for $k = 1$ and $k = 2$ are shown in Figs. \ref{Bipartite1} and \ref{Bipartite2}. Our results obtained from the solution to the ILP show that the 5 transformers can be monitored with 4 sensors when $k = 1$, and 3 sensors when $k = 2$. As shown in Fig. \ref{Bipartite1}, for $k = 1$, if 4 sensors are deployed at nodes 14, 19, 27, and 30 (equivalently 4 colors A, B, C, and D are injected at these nodes, shown in Fig. \ref{Bipartite1} by A*, B*, C*, and D*), the 5 transformers T1 through T5 will receive \emph{unique} colors AC, A, B, CD, and D, respectively. Similarly, for $k = 2$, if 3 sensors are deployed at nodes 8, 27, and 35 (equivalently 3 colors A, B, and C are injected at these nodes, shown in Fig. \ref{Bipartite2} by A*, B*, and C*), the 5 transformers T1 through T5 will receive \emph{unique} colors AB, ABC, A, B, and BC respectively.   

The significance of each transformer receiving a unique color (or a unique signature), is the following. In the example shown in Fig. \ref{Bipartite2}, if colors A, B and C are injected at nodes 8, 27 and 35 (i.e., PMUs A, B, and C are placed at these locations, among the 40 (6-45) potential locations), the transformers T1-T5 will receive colors AB, ABC, A, B, and BC, respectively. Suppose that the control room has three indicator lamps, 1, 2, and 3, corresponding to PMUs A, B, and C, respectively. As long as the width of the SNR ratio is within the normal range, the lamps are green. As soon as the width of the SNR ratio exceeds the normal range, the corresponding lamps turn red. An operator, at the control room, can interpret the status of the five transformers, in the following way: (i) The transformer T1 is failing if only lamps 1 and 2 turn red, (ii) T2 is failing if lamps 1, 2 and 3 turn red, (iii) T3 is failing if lamp 1 turns red, and so on.  

\begin{table}[h!]
\centering
\caption{No. of sensors needed in IEEE, PEGASE, and Polish systems for $k = 1, 2$.}
\begin{tabular}{ |c|c|c|c| }
 \hline
 Bus System & No. of Transformers & \multicolumn{2}{c|}{No. of Sensors} \\ \cline{3-4}
 &  & $k = 1$  & $k = 2$ \\ \hline
IEEE 14 & 5 & 4 & 3 \\ \hline
IEEE 30 & 7 & 6 & 4 \\ \hline
IEEE 57 & 14 & 13 & 10 \\ \hline
PEGASE 89 & 10 & 10 & 6\\ \hline
IEEE 118 & 9 &9 & 5 \\ \hline
Polish 2383 & 155 & 155 & 106 \\ \hline
\end{tabular}
\label{Colors}
\vspace{6.00pt}
\vspace{-8.0mm}
\end{table}

Our results for power system test cases are tabulated in Table \ref{Colors}. The results show that the number of sensors needed to monitor all the transformers are fewer than the number of transformers. On an average there were 6.90\% and 37.90\% savings in the number of sensors using our technique for $k = 1$ and $k = 2$, respectively. From Fig. \ref{distance}, it can be seen that the difference in the width of the SNR band in dB at substations $S_1$ and $S_2$ (1 and 2 hop distance away respectively, from the transformer) is minimal.  Accordingly, we can use $k = 2$ results, which implies that  significant savings (37.90\%) can be realized using our technique. The ILPs for the test cases were computed using GUROBI for python. An Intel Core i5-6300HQ CPU with 2.30 GHz and 32 GB RAM was used for our experiments. The computation time varied from 0.17 seconds, for the smallest test case ($|V_1|$ = 5, $|V_2|$ = 40, $|E|$ = 36, $k$ = 1), to 25.18 seconds ($|V_1|$ = 155, $|V_2|$ = 5,772, $|E|$ = 3,655, $k$ = 2) for the largest. As the computation times for these test cases were only a few seconds, we expect that for larger systems involving thousands of buses and hundreds of transformers, the problem can still be solved within a short period of time. 
\vspace{-8.00pt}
\section{Conclusion}
\label{Conclusions}
We present a novel technique involving PMU-based metrics and Identifying Code to find the least number of sensors to monitor the health of the critical equipments, such as HV power transformers. In the future, we plan to investigate (i) a fault tolerant monitoring system, where the system will be able to \emph{uniquely} identify a failing critical equipment, even when one or more of the sensors are malfunctioning, and (ii) multiple simultaneous failure of critical equipments, in the sense that, not only failure of individual equipments will have a \emph{unique} signature, but also failure of a set of equipments will have a \emph{unique fault signature}. 
\bibliographystyle{ieeetr}

\begin{thebibliography}
\footnotesize
\bibitem{R38} V. Salehi, A. Mohamed, A. Mazloomzadeh, and O. A. Mohammed,``Laboratory-based smart power system, part II: control, monitoring, and protection,'' {\em IEEE Trans. Smart Grid}, vol. 3, no. 3, Sep. 2012.










\bibitem{R10} A. Pal, A. K. S. Vullikanti, and S. S. Ravi, ``A PMU placement scheme considering realistic costs and modern trends in relaying,'' {\em IEEE Trans. Power Syst.}, vol. 32, no. 1, Jan. 2017.

\bibitem{R11} A. Pal, C. Mishra, A. K. S. Vullikanti, and S. S. Ravi, ``General optimal substation coverage algorithm for phasor measurement unit placement in practical systems,'' {\em IET Gener., Transm. Distrib.}, vol. 11, no. 2, Jan. 2017.


















\bibitem{R24} M. G. Karpovsky, K. Chakrabarty, and L. B. Levitin, ``On a New Class of Codes for Identifying Vertices in Graphs,'' {\em IEEE Trans. Inf. Theory}, vol. 44, no. 2, Mar. 1998.

\bibitem{R25} M. Laifenfeld and A. Trachtenberg, ``Identifying Codes and Covering Problems,'' {\em IEEE Trans. Inf. Theory}, vol. 54, no. 9, Sep. 2008.








\bibitem{Discriminating} E. Charbit, I. Charon, G. Cohen, and O. Hudry, ``Discriminating Codes in Bipartite Graphs,'' {\em Electronic Notes in Discrete Mathematics}, vol. 26, Sep. 2006.

\bibitem{R16} ``Transformer explodes into fireball at SRP substation in Avondale.'' [Online]. Available: \url{http://www.azfamily.com/story/32111291/transformer-explodes-into-fireball-at-srp-substation-in-avondale}

\end{thebibliography}

\end{document}